\documentclass{article}
\usepackage{amsmath}
\usepackage{amssymb}
\usepackage{amsfonts}

\input{tcilatex}
\begin{document}

\author{J. G. Cardoso\thanks{%
jorge.cardoso@udesc.br}\thanks{%
Retired from the Santa Catarina State University.} \\
Department of Mathematics\\
Centre for Technological Sciences-UDESC\\
89223-100 Joinville Santa Catarina\\
Brazil.}
\title{A Relativistic Quantum Mechanical Formulation of Wu-Shaknov
Entanglements}
\date{ }
\maketitle

\begin{abstract}
It is suggested that a spontaneous formation of quantum mechanical
entanglements does actually occur between spin, helicity and polarization
states involved in creation-annihilation processes. A framework built up
earlier that presumably affords a unified relativistic quantum mechanical
description of free particle-antiparticle pairs of any spin is used to
propose a comprehensive formulation of Wu-Shaknov entanglements. Explicit
relationships between amplitudes and Born-like probabilities associated to
incoming and outgoing states are obtained thereafter on the basis of certain
theoretical assumptions. The behaviours of these relationships under the
action of a dynamical group which is taken to supply observational
correlations for different frames in spacetime are then specified. It
appears that a modified version of some of the conventional conservation
laws may afford a set of general selection rules for the formation of
relativistic entanglements.
\end{abstract}

\section{Introduction}

Quantum entanglements were reliably observed for the first time [1] by Wu
and Shaknov (WS) in 1950. Their experimental set up produced locally
entangled photons coming from the annihilation of electron-positron pairs,
and thereby confirmed the physical legitimacy of a prediction that had been
made by Dirac [2] of an occurrent correlation between the polarizations of
outgoing photons in annihilation processes. The original concept of quantum
mechanical entanglements had nevertheless been sorted out much earlier by
Schr\"{o}dinger [3,4], who suggested that entanglements could be undone as
amplitudes of quantum states must undergo well defined changes whenever the
corresponding systems are spatially separated. Several observations of the
occurrence of a broader class of WS-like entanglements have been reported in
recent years, with all of them having arisen from collider-based studies
which probe an entanglement phenomenology in creation-annihilation processes
at the highest accessible experimental energy scales available nowadays\
(see, for instance, Refs. [5,6]).

The present work suggests that a spontaneous formation of quantum mechanical
entanglements does actually occur between spin, helicity and polarization
states involved in creation-annihilation processes. A framework built up in
Ref. [7], which presumably affords a unified relativistic quantum mechanical
description of free particle-antiparticle pairs of any spin, is utilized to
propose a comprehensive formulation of entanglements occurring in processes
of that kind. The construction of this framework was crucially based upon
the combination of the existence [8] of a four-to-one homomorphism between $%
SU(2,2)$ and the orthocronous proper conformal group $\mathcal{C}%
_{+}^{\uparrow }$ of special relativity with the statement that $\mathcal{C}%
_{+}^{\uparrow }$ comprises the maximal spacetime extension of the Poincar%
\'{e} group [9]. It has supported the view whereby the physical inner
structure of the Schr\"{o}dinger quantum mechanical picture may be entirely
fitted in with the realm of special relativity, while not only maintaining
the traditional concept of Schr\"{o}dinger quantum mechanical states along
with the old interpretation of Born probabilities, but also supplying
relativistic arrangements of observational correlations for dynamical
variables and measurements of states for free fermions and bosons of any
rest mass.

We will see that explicit relationships between amplitudes and Born
probabilities associated locally to entangled incoming and outgoing states
come about on the basis of certain theoretical assumptions. Indeed, every
process shall be taken to admit a linear mapping which gives rise to the
derivation of the statistical relationships for the process. The framework
thus specifies the behaviours of these relationships under the action of a
dynamical group which had been effectively taken in Refs. [7,10] to supply
observational correlations for different observers in spacetime. It appears
that a modified version of some of the conventional conservation laws may
afford a set of general selection rules for the formation of relativistic
entanglements. In annihilation processes, the formation of entanglements
between the incoming particles should happen within finite proper-time
intervals, just prior to the emergence of the relevant outgoing particles.
However, the incoming particles of such processes must constitute composite
systems having covariant states before the occurrence of entanglements, but
their states eventually become reduced in conjunction with the entanglement
formation. Moreover, entanglements occurring in creation pairs, as well as
the ones occurring in multiple-particle outcomes of annihilation processes,
are already formed at the beginning of the emergence of the outgoing
particles but, in any case, the intrinsic statistical character of the
processes must be thought of as conforming to the dynamical role that is
broadly ascribed to conservation laws.

In order to fulfill the main aims of the work and to enhance our methods to
a sufficient extent, it will be enough in principle to consider processes
where either an incoming boson or antiboson creates a fermion-antifermion
pair, and annihilation processes whose outgoing particles show up as pairs
of bosons or pairs of antibosons (WS processes). The labelling and the
"reflected" Dirac-like bra-ket notations used in Ref. [7], as well as the
spacetime signature $(+---)$, shall all be adopted throughout what follows.
A few further conventions will be explained occasionally. In Section 2, we
exhibit some mechanical preliminaries. The processes are considered
separately in Sections 3 and 4. In Section 5, we bring out the observational
correlations of interest and, in Section 6, some physical remarks on the
overall work are drawn up.

\section{Preliminaries}

Such as in Ref. [7], the spaces of states for any particle and antiparticle
that make up a free particle-antiparticle pair $(p^{+},p^{-})$ moving in a
given spacetime frame are, respectively, identified with the slices $%
\mathfrak{C}^{+}(p^{+})$ and $\mathfrak{C}^{-}(p^{-})$ carried by a copy $%
\mathfrak{C}(p^{+},p^{-})$ of the representation space of $\mathcal{C}%
_{+}^{\uparrow }$ through the orthogonal sum%
\begin{equation}
\mathfrak{C}(p^{+},p^{-})=\mathfrak{C}^{+}(p^{+})\dotplus \mathfrak{C}%
^{-}(p^{-}),  \label{1}
\end{equation}%
with $\mathfrak{C}^{\pm }(p^{\pm })\simeq 
\mathbb{C}
^{2}.$ All dynamical variables defined thereabout are taken to carry a
symbolic coordinate-free operator restriction character so as to operate
linearly on spaces of states independently of the action of any generator of 
$\mathcal{C}_{+}^{\uparrow }.$ This definition implies that the conservation
laws which govern creation-annihilation processes may be expressed simply as
equalities between sums of spectral entries.

Upon carrying out the formulation of the processes to be allowed for herein,
we shall choose to consider just helicity states for spin $\pm 1/2$
fermion-antifermion pairs and for bosons of polarizations $\pm 1$. In
essence, the viability of this choice stems from the fact that there exist
local classes of linear operators on $\mathfrak{C}(p^{+},p^{-})$ which can
be employed for changing to helicity states both the spin states of fermions
and the polarization states of bosons.\footnote{%
A scheme for changing in a spacetime frame the degrees of freedom of any
particle-antiparticle pair is exhibited in the last four paragraphs of the
third Section of Ref. [7].}

The canonical basis%
\begin{equation}
\{<e_{(\mu )}(p^{+},p^{-})\mid \}=\{<e_{(0)}^{(p+)}\mid ,<e_{(1)}^{(p+)}\mid
,<e_{(0)}^{(p-)}\mid ,<e_{(1)}^{(p-)}\mid \}\text{ }  \label{1lin}
\end{equation}%
of $\mathfrak{C}(p^{+},p^{-})$ serves as a prototype of a complete set of
eigenstates for the up-down spin and horizontal-vertical polarization
restrictions of $(p^{+},p^{-}).$ Any helicity restrictions $h^{(p\pm )}$ for 
$p^{\pm }$ are defined together with an assignment of Hilbert-unitary
operators like%
\begin{equation}
\text{\textsc{H}}(p^{+},p^{-})=\left( 
\begin{array}{ll}
\text{\textsc{H}}^{(p+)} & 0 \\ 
0 & \text{\textsc{H}}^{(p-)}%
\end{array}%
\right) ,  \label{op1}
\end{equation}%
whose matrix representation is subjected to the locality condition%
\begin{equation}
\text{any representative matrix for \textsc{H}}(p^{+},p^{-})\notin
SU(2,2)\cap U(4).  \label{2}
\end{equation}%
We have the basis changes%
\begin{equation}
<\mathfrak{h}_{(\mu )}^{(p\pm )}\mid =<e_{(\mu )}^{(p\pm )}\mid \text{%
\textsc{H}}^{(p\pm )},  \label{3}
\end{equation}%
which supply us with the left-right states%
\begin{equation}
\{<\mathfrak{h}_{(0)}^{(p\pm )}\mid ,<\mathfrak{h}_{(1)}^{(p\pm )}\mid \}
\label{4}
\end{equation}%
along with the helicity Gram restrictions\footnote{%
Here as elsewhere, $g^{\pm }$ denotes the fundamental Gram restrictions for
any copy of the representation space. The superscript "dagger" has been
taken to denote the Hermitian conjugation for the Hilbert inner product.}%
\begin{equation}
\mathfrak{H}^{(\text{\textsc{H}}p\pm )}=\text{\textsc{H}}^{(p\pm )\dag
}g^{\pm }\text{\textsc{H}}^{(p\pm )},  \label{5Lin}
\end{equation}%
whilst satisfying the unitarity relation%
\begin{equation}
<\mathfrak{h}_{(\mu )}^{(p\pm )}\mid \mathfrak{h}_{(\nu )}^{(p\pm )}>_{%
\mathfrak{H}^{(\text{\textsc{H}}p\pm )}}=<e_{(\mu )}^{(p\pm )}\mid e_{(\nu
)}^{(p\pm )}>_{g^{\pm }}  \label{6Lin}
\end{equation}%
and the orthogonality condition%
\begin{equation}
<\mathfrak{h}_{(\mu )}^{(p\pm )}\mid \mathfrak{H}^{(\text{\textsc{H}}p\pm
)}\parallel \mathfrak{h}_{(\nu )}^{(p\pm )}>_{\mathfrak{H}^{(\text{\textsc{H}%
}p\pm )}}=\Delta _{\mu \nu }^{\pm },  \label{cond1}
\end{equation}%
where%
\begin{equation}
(\Delta _{\mu \nu }^{\pm })=\left( 
\begin{array}{ll}
1 & 0 \\ 
0 & 1%
\end{array}%
\right) .  \label{cond2}
\end{equation}

For the case of any massive (massless) pair, the definition (\ref{op1})
absorbs implicitly some timelike (null) directions in a spacetime frame as
the locally specified directions of motion of $p^{+}$ and $p^{-}$. Changing
local spacetime directions, produces the requirement for selecting other 
\textsc{H}-operators but amplitudes of states, and thence Born probabilities
too, must be left unaltered when any devices like (\ref{3}) are accounted
for (for further details, see Eq. (3.62) of Ref. [7]).

For a fermionic pair $(f^{+},f^{-})$, Eq. (\ref{3}) gives the states%
\begin{equation}
\{<\mathfrak{h}_{(0)}^{(f\pm )}\mid ,<\mathfrak{h}_{(1)}^{(f\pm )}\mid
\}\subset \mathfrak{C}^{\pm }(f^{\pm }),  \label{add6}
\end{equation}%
which yield the helicity-spin spectral entries%
\begin{equation}
h_{\mu \nu }^{(f\pm )}=<\mathfrak{h}_{(\mu )}^{(f\pm )}\mid h^{(f\pm
)}\parallel \mathfrak{h}_{(\nu )}^{(f\pm )}>_{\mathfrak{H}^{(\text{\textsc{H}%
}f\pm )}}=\Sigma _{\mu \nu }^{(f\pm )}  \label{5}
\end{equation}%
and%
\begin{equation}
\Sigma _{\mu \nu }^{(f\pm )}=<e_{(\mu )}^{(f\pm )}\mid \Sigma ^{(f\pm
)}\parallel e_{(\nu )}^{(f\pm )}>_{g^{\pm }},  \label{add5}
\end{equation}%
provided that%
\begin{equation}
h^{(f\pm )}=\text{\textsc{H}}^{(f\pm )\dag }\Sigma ^{(f\pm )}\text{\textsc{H}%
}^{(f\pm )}  \label{6}
\end{equation}%
and%
\begin{equation}
\mathfrak{H}^{(\text{\textsc{H}}f\pm )}=\text{\textsc{H}}^{(f\pm )\dag
}g^{\pm }\text{\textsc{H}}^{(f\pm )},  \label{add1}
\end{equation}%
with $\Sigma ^{(f\pm )}$ accordingly standing for the spin operators of $%
f^{\pm },$ and $\mathfrak{H}^{(f\pm )}$ amounting to the respective Gram
restrictions. Equations (\ref{5}) and (\ref{add5}) thus give the spectra%
\begin{equation}
(h_{\mu \nu }^{(f\pm )})=%
\begin{pmatrix}
\pm \frac{1}{2} & 0 \\ 
0 & \mp \frac{1}{2}%
\end{pmatrix}%
=\Sigma _{\mu \nu }^{(f\pm )}.  \label{spec1}
\end{equation}

Similarly, for a bosonic pair $(b^{+},b^{-})$ with polarization operators $%
\Pi ^{(b\pm )}$, we have the matrix elements%
\begin{equation}
h_{\mu \nu }^{(b\pm )}=<\mathfrak{h}_{(\mu )}^{(b\pm )}\mid h^{(b\pm
)}\parallel \mathfrak{h}_{(\nu )}^{(b\pm )}>_{\mathfrak{H}^{(\text{\textsc{H}%
}b\pm )}}=\Pi _{\mu \nu }^{(b\pm )}  \label{7}
\end{equation}%
and%
\begin{equation}
\Pi _{\mu \nu }^{(b\pm )}=<e_{(\mu )}^{(b\pm )}\mid \Pi ^{(b\pm )}\parallel
e_{(\nu )}^{(b\pm )}>_{g^{\pm }},  \label{7Lin}
\end{equation}%
together with the prescriptions%
\begin{equation}
h^{(b\pm )}=\text{\textsc{H}}^{(b\pm )\dag }\Pi ^{(b\pm )}\text{\textsc{H}}%
^{(b\pm )}  \label{8}
\end{equation}%
and%
\begin{equation}
\mathfrak{H}^{(\text{\textsc{H}}b\pm )}=\text{\textsc{H}}^{(b\pm )\dag
}g^{\pm }\text{\textsc{H}}^{(b\pm )},  \label{9}
\end{equation}%
as well as with the inclusion%
\begin{equation}
\{<\mathfrak{h}_{(0)}^{(b\pm )}\mid ,<\mathfrak{h}_{(1)}^{(b\pm )}\mid
\}\subset \mathfrak{C}^{\pm }(b^{\pm })  \label{15}
\end{equation}%
and the spectra%
\begin{equation}
(h_{\mu \nu }^{(b\pm )})=%
\begin{pmatrix}
\pm 1 & 0 \\ 
0 & \mp 1%
\end{pmatrix}%
=(\Pi _{\mu \nu }^{(b\pm )}).  \label{spec1lin}
\end{equation}%
In the forthcoming Sections, we will often make use of the simplified
notation%
\begin{equation}
<\mathfrak{h}_{(0)}^{(p\pm )}\mid =<0^{(p\pm )}\mid ,\text{ }<\mathfrak{h}%
_{(1)}^{(p\pm )}\mid =<1^{(p\pm )}\mid .  \label{add19}
\end{equation}

\section{Creation Processes}

We shall now consider a process in which either an incoming boson $B^{+}$ or
antiboson $B^{-}$ creates a fermion-antifermion pair $(F^{+},F^{-})$. For $%
B^{+}$, say, the incoming state for the process is written as the
superposition 
\begin{equation}
<B^{+}\mid =B_{+}^{\mu }<\mathfrak{h}_{(\mu )}^{(B+)}\mid ,  \label{16}
\end{equation}%
which obviously belongs to $\mathfrak{C}^{+}(B^{+}).$ Its amplitudes are, by
(\ref{cond1}), given as%
\begin{equation}
B_{+}^{0}=<B^{+}\mid \mathfrak{H}^{(\text{\textsc{HB}}+)}\parallel
0^{(B+)}>_{\mathfrak{H}^{(\text{\textsc{HB}}+)}}  \label{a1}
\end{equation}%
and%
\begin{equation}
B_{+}^{1}=<B^{+}\mid \mathfrak{H}^{(\text{\textsc{HB}}+)}\parallel
1^{(B+)}>_{\mathfrak{H}^{(\text{\textsc{HB}}+)}},  \label{a2}
\end{equation}%
which satisfy the normalization condition%
\begin{equation}
<B^{+}\mid \mathfrak{H}^{(\text{\textsc{HB}}+)}\parallel B^{+}>_{\mathfrak{H}%
^{(\text{\textsc{HB}}+)}}=\mid B_{+}^{0}\mid ^{2}+\mid B_{+}^{1}\mid ^{2}=1.
\label{a3}
\end{equation}

Since the created particles of either case supposedly emerge in entangled
states having specific statistical weights and obeying characteristic
conservation laws, we must regard the corresponding outgoing state as a
non-covariant reduced superposition like%
\begin{equation}
<\text{ent }F^{+}F^{-}\mid =c_{+-}^{00}<0^{(F+)}0^{(F-)}\mid +\text{ }%
c_{+-}^{11}<1^{(F+)}1^{(F-)}\mid ,  \label{17}
\end{equation}%
with the general outer-product notation%
\begin{equation}
<\mathfrak{h}_{(\mu )}^{(p\pm )}\mid \otimes <\mathfrak{h}_{(\nu )}^{(p\pm
)}\mid =<\mathfrak{h}_{(\mu )}^{(p\pm )}\mathfrak{h}_{(\nu )}^{(p\pm )}\mid 
\label{30}
\end{equation}%
having been adopted for spelling out (\ref{17}). It is clear that $<$ent $%
F^{+}F^{-}\mid $ belongs to the outer-product space%
\begin{equation}
\mathfrak{C}^{+}(F^{+})\otimes \mathfrak{C}^{-}(F^{-}),  \label{99}
\end{equation}%
which carries the Gram operator%
\begin{equation}
\mathfrak{H}^{(\text{\textsc{HF}}^{+}\text{\textsc{F}}^{-})}=\mathfrak{H}^{(%
\text{\textsc{HF}}+)}\otimes \mathfrak{H}^{(\text{\textsc{HF}}-)}.
\label{91}
\end{equation}%
Hence, the amplitudes of the state (\ref{17}) are given by%
\begin{equation}
c_{+-}^{00}=<\text{ent }F^{+}F^{-}\mid \mathfrak{H}^{(\text{\textsc{HF}}^{+}%
\text{\textsc{F}}^{-})}\parallel 0^{(F+)}0^{(F-)}>_{\mathfrak{H}^{(\text{%
\textsc{HF}}^{+}\text{\textsc{F}}^{-})}}  \label{21}
\end{equation}%
and%
\begin{equation}
c_{+-}^{11}=<\text{ent }F^{+}F^{-}\mid \mathfrak{H}^{(\text{\textsc{HF}}^{+}%
\text{\textsc{F}}^{-})}\parallel 1^{(F+)}1^{(F-)}>_{\mathfrak{H}^{(\text{%
\textsc{HF}}^{+}\text{\textsc{F}}^{-})}},  \label{23}
\end{equation}%
in agreement with the condition%
\begin{equation}
<\text{ent }F^{+}F^{-}\mid \mathfrak{H}^{(\text{\textsc{HF}}^{+}\text{%
\textsc{F}}^{-})}\parallel \text{ent }F^{+}F^{-}>_{\mathfrak{H}^{(\text{%
\textsc{HF}}^{+}\text{\textsc{F}}^{-})}}=1.  \label{92}
\end{equation}

Some relations between the amplitudes and Born probabilities associated to
the states (\ref{16}) and (\ref{17}) can be attained by first defining a
creation mapping%
\begin{equation}
\mathcal{C}^{+}(B^{+}):\mathfrak{C}^{+}(B^{+})\longrightarrow \mathfrak{C}%
^{+}(F^{+})\otimes \mathfrak{C}^{-}(F^{-}),  \label{93}
\end{equation}%
which presumptively bears linearity and invertibility. We have the
correspondence%
\begin{equation}
<B^{+}\mid \mathcal{\longmapsto }<\text{ent }F^{+}F^{-}\mid =<B^{+}\mid 
\mathcal{C}^{+}(B^{+}),  \label{94}
\end{equation}%
whence we can likewise write%
\begin{equation}
<\text{ent }F^{+}F^{-}\mid \text{ent }F^{+}F^{-}>_{\mathfrak{H}^{(\text{%
\textsc{HF}}^{+}\text{\textsc{F}}^{-})}}=<B^{+}\mid \mathcal{C}^{+}(B^{+})%
\mathcal{C}^{+\#}(B^{+})\parallel B^{+}>_{\mathfrak{H}^{(\text{\textsc{HB}}%
+)}},  \label{in}
\end{equation}%
with the symbol "$\#$" thus denoting the Hermitian conjugation with respect
to the $\mathfrak{H}^{(\text{\textsc{HB}}+)}$-inner product. Then, by
assuming that%
\begin{equation}
B_{+}^{0}<0^{(B+)}\mid \mathcal{C}^{+}(B^{+})=c_{+-}^{00}<0^{(F+)}0^{(F-)}%
\mid  \label{prob1}
\end{equation}%
and%
\begin{equation}
B_{+}^{1}<1^{(B+)}\mid \mathcal{C}^{+}(B^{+})=c_{+-}^{11}<1^{(F+)}1^{(F-)}%
\mid ,  \label{prob2}
\end{equation}%
we obtain the left-right probabilities%
\begin{equation}
\mid \text{ }c_{+-}^{00}\mid ^{2}=w_{L}\mid B_{+}^{0}\mid ^{2}  \label{prob3}
\end{equation}%
and%
\begin{equation}
\mid \text{ }c_{+-}^{11}\mid ^{2}=w_{R}\mid B_{+}^{1}\mid ^{2},
\label{prob4}
\end{equation}%
together with the creation weights%
\begin{equation}
w_{L}=\mid <0^{(B+)}\mid \mathcal{C}^{+}(B^{+})\mathcal{C}%
^{+\#}(B^{+})\parallel 0^{(B+)}>_{\mathfrak{H}^{(\text{\textsc{HB}}+)}}\mid
^{2}  \label{p1}
\end{equation}%
and%
\begin{equation}
w_{R}=\mid <1^{(B+)}\mid \mathcal{C}^{+}(B^{+})\mathcal{C}%
^{+\#}(B^{+})\parallel 1^{(B+)}>_{\mathfrak{H}^{(\text{\textsc{HB}}+)}}\mid
^{2}.  \label{p2}
\end{equation}

\section{Annihilation Processes}

By this point, we consider a typical WS process having a couple of bosons $%
b^{+}$ and $B^{+}$ as its outcome. In accordance with was brought up in
Section 1, we initially regard the incoming particles as a composite system
having the covariant fermionic state%
\begin{equation}
<f^{+}f^{-}\mid =f_{+}^{\mu }f_{-}^{\nu }<\mathfrak{h}_{(\mu )}^{(f+)}\mid
\otimes <\mathfrak{h}_{(\nu )}^{(f-)}\mid ,  \label{proc1}
\end{equation}%
which of course belongs to a product space like (\ref{99}), in addition to
obeying the normalization condition%
\begin{equation}
<f^{+}f^{-}\mid \mathfrak{H}^{(\text{\textsc{H}f}^{+}\text{f}^{-})}\parallel
f^{+}f^{-}>_{\mathfrak{H}^{(\text{\textsc{H}f}^{+}\text{f}^{-})}}=1.
\label{proc9}
\end{equation}%
The formation of an entangled state $<$ent $f^{+}f^{-}\mid $ between $f^{+}$
and $f^{-}$ is accomplished from the formal definition%
\begin{equation}
\varepsilon (f^{+},f^{-}):\mathfrak{C}^{+}(f^{+})\otimes \mathfrak{C}%
^{-}(f^{-})\longrightarrow \mathfrak{C}^{+}(f^{+})\otimes \mathfrak{C}%
^{-}(f^{-}),  \label{proc2}
\end{equation}%
where $\varepsilon (f^{+},f^{-})$ is a linear operator which must yield the
normalization-preserving correspondence%
\begin{equation}
\varepsilon (f^{+},f^{-}):<f^{+}f^{-}\mid \longmapsto <\text{ent }%
f^{+}f^{-}\mid .  \label{ent1}
\end{equation}

Because of the requirement coming from conservation laws, which imposes that
the spectral entries for the helicities of the incoming and outgoing
particles should be suitably balanced, the outgoing state for the process
must right away be taken uniquely as a normalized entangled two-boson state
like%
\begin{equation}
<\text{ent }b^{+}B^{+}\mid =\beta _{++}^{01}<0^{(b+)}1^{(B+)}\mid +\text{ }%
\beta _{++}^{10}<1^{(b+)}0^{(B+)}\mid ,  \label{proc7}
\end{equation}%
whereas the entanglement reduction of (\ref{proc1}) via (\ref{ent1}) must
therefore yield a state like%
\begin{equation}
<\text{ent }f^{+}f^{-}\mid =a_{+-}^{01}<0^{(f+)}1^{(f-)}\mid +\text{ }%
a_{+-}^{10}<1^{(f+)}0^{(f-)}\mid ,  \label{proc4}
\end{equation}%
with $a_{+-}^{01}$ and $a_{+-}^{10},$ for instance, being expressed in much
the same way as (\ref{21}) and (\ref{23}), namely,%
\begin{equation}
a_{+-}^{01}=<\text{ent }f^{+}f^{-}\mid \mathfrak{H}^{(\text{\textsc{H}f}^{+}%
\text{f}^{-})}\parallel 0^{(f+)}1^{(f-)}>_{\mathfrak{H}^{(\text{\textsc{H}f}%
^{+}\text{f}^{-})}}  \label{proc5}
\end{equation}%
and%
\begin{equation}
a_{+-}^{10}=<\text{ent }f^{+}f^{-}\mid \mathfrak{H}^{(\text{\textsc{H}f}^{+}%
\text{f}^{-})}\parallel 1^{(f+)}0^{(f-)}>_{\mathfrak{H}^{(\text{\textsc{H}f}%
^{+}\text{f}^{-})}}.  \label{proc6}
\end{equation}

We assume that the statistical relationships for the process are afforded by
the action of an annihilation mapping%
\begin{equation}
\mathfrak{A}(f^{+}f^{-}):\mathfrak{C}^{+}(f^{+})\otimes \mathfrak{C}%
^{-}(f^{-})\longrightarrow \mathfrak{C}^{+}(b^{+})\otimes \mathfrak{C}%
^{+}(B^{+}),  \label{proc21}
\end{equation}%
which then takes $<$ent $f^{+}f^{-}\mid $ into $<$ent $b^{+}B^{+}\mid $
linearly. Based upon this assumption, we also allow for the combined patterns%
\begin{equation}
a_{+-}^{01}<0^{(f+)}1^{(f-)}\mid \mathfrak{A}(f^{+}f^{-})=\beta
_{++}^{01}<0^{(b+)}1^{(B+)}\mid  \label{proc15}
\end{equation}%
and%
\begin{equation}
a_{+-}^{10}<1^{(f+)}0^{(f-)}\mid \mathfrak{A}(f^{+}f^{-})=\beta
_{++}^{10}<1^{(b+)}0^{(B+)}\mid ,  \label{proc16}
\end{equation}%
which consequently enable us to write down the probabilities%
\begin{equation}
\mid \beta _{++}^{01}\mid ^{2}=w_{01}\mid a_{+-}^{01}\mid ^{2}  \label{pr1}
\end{equation}%
and%
\begin{equation}
\mid \beta _{++}^{10}\mid ^{2}=w_{10}\mid a_{+-}^{10}\mid ^{2},  \label{p3}
\end{equation}%
along with the annihilation weights%
\begin{equation}
w_{01}=\mid <0^{(f+)}1^{(f-)}\mid \mathfrak{A}(f^{+}f^{-})\mathfrak{A}%
^{\blacklozenge }(f^{+}f^{-})\parallel 0^{(f+)}1^{(f-)}>_{\mathfrak{H}^{(%
\text{\textsc{H}f}^{+}\text{f}^{-})}}\mid ^{2}  \label{p4}
\end{equation}%
and\footnote{%
We suggest that the duration of entanglements occurring in annihilation
processes which involve two outgoing particles is of the same order as that
of the proper mean lifetime in vacuum of the ground state of positronium
[11], namely, 0.125 $\times $ 10$^{\text{-9}}$ s.}%
\begin{equation}
w_{10}=\mid <1^{(f+)}0^{(f-)}\mid \mathfrak{A}(f^{+}f^{-})\mathfrak{A}%
^{\blacklozenge }(f^{+}f^{-})\parallel 1^{(f+)}0^{(f-)}>_{\mathfrak{H}^{(%
\text{\textsc{H}f}^{+}\text{f}^{-})}}\mid ^{2},  \label{p5}
\end{equation}%
where the symbol "$\blacklozenge $" stands for the Hermitian conjugation
with respect to $\mathfrak{H}^{(\text{\textsc{H}f}^{+}\text{f}^{-})}.$

\section{Observational correlations}

Within the framework of Ref. [7], any results of dynamical observations
performed in different spacetime frames, are correlated by the
operationality of the unitary intersection%
\begin{equation}
SU(4)\supset \mathcal{P}_{Dyn}^{+\uparrow }=\mathcal{P}_{+}^{\uparrow }\cap
U(4),  \label{frame1}
\end{equation}%
where $\mathcal{P}_{+}^{\uparrow }$ amounts to the orthochronous proper
Poincar\'{e} subgroup of an adequate realization of $SU(2,2).$ An arbitrary
element of $\mathcal{P}_{Dyn}^{+\uparrow }$ possesses the form%
\begin{equation}
(u_{\mu \nu }(Dyn))=\frac{1}{\sqrt{2}}\left( 
\begin{array}{cc}
(I_{2}+iW)\beta & 0_{2} \\ 
0_{2} & \beta ^{\dag }(I_{2}-iW)%
\end{array}%
\right) ,  \label{frame2}
\end{equation}%
with $\beta \in U(2)$ representing boosts along some spacetime directions or
otherwise proper rotations as for $\beta \in SU(2)$, and $W$ being the van
der Waerden $(2\times 2)$-matrix associated to a time-like or space-like
Minkowskian translation normalized as%
\begin{equation}
W^{2}=\text{identity }(2\times 2)\text{-matrix.}  \label{f3}
\end{equation}%
Changes of spacetime frames are then induced by the action of $\mathcal{P}%
_{Dyn}^{+\uparrow },$ whereas the array (\ref{frame2}) may be looked upon as
the representative matrix in an admissible basis of a pseudo-unitary
operator $u_{Dyn}$ that acts on some $\mathfrak{C}(p^{+},p^{-}).$ We thus
have\footnote{%
Only bases related to one another by $SU(2,2)$-transformations bear
admissibility. We should stress that the form of (\ref{frame2}) preserves
the individual particle-antiparticle characters of any dynamical
descriptions.}%
\begin{equation}
u_{Dyn}{}=\left( 
\begin{array}{ll}
u_{Dyn}^{+}{} & 0 \\ 
0 & u_{Dyn}^{-}{}%
\end{array}%
\right) ,\text{ }u_{Dyn}^{\pm \bigstar }{}=(u_{Dyn}^{\pm
}{})^{-1}=u_{Dyn}^{\pm \dag }{},  \label{f4}
\end{equation}%
with the symbol "$\bigstar $" denoting the pseudo-Hermitian conjugation with
respect to the $g^{\pm }$-inner products, and%
\begin{equation}
<e_{(\mu )}^{\prime (p\pm )}\mid =<e_{(\mu )}^{(p\pm )}\mid u_{Dyn}^{\pm }{}.
\label{f5}
\end{equation}

The decomposition of $\mathfrak{C}(p^{+},p^{-})$ as prescribed by Eq. (\ref%
{1}) must be regarded as an invariant feature of the processes. Setting%
\begin{equation}
\text{\textsc{H}}^{\prime (p\pm )}=u_{Dyn}^{\pm \bigstar }\text{\textsc{H}}%
^{(p\pm )},  \label{f6}
\end{equation}%
entails the invariance of the states (\ref{4}) which, in turn, yields the
invariance of Eq. (\ref{6Lin}). In effect, we have the development%
\begin{eqnarray}
&<&\mathfrak{h}_{(\mu )}^{\prime (p\pm )}\mid \mathfrak{h}_{(\nu )}^{\prime
(p\pm )}>_{\mathfrak{H}^{(\text{\textsc{H}}^{\prime }p\pm )}}=<e_{(\mu
)}^{(p\pm )}\mid u_{Dyn}^{\pm }\text{\textsc{H}}^{\prime (p\pm )}\mid
u_{Dyn}^{\pm }\text{\textsc{H}}^{\prime (p\pm )}\mid e_{(\nu )}^{(p\pm )}>_{%
\mathfrak{H}^{(\text{\textsc{H}}^{\prime }p\pm )}}  \notag \\
&=&<e_{(\mu )}^{(p\pm )}\mid \text{\textsc{H}}^{(p\pm )}\mid \text{\textsc{H}%
}^{(p\pm )}\mid e_{(\nu )}^{(p\pm )}>_{\mathfrak{H}^{(\text{\textsc{H}}p\pm
)}}=<\mathfrak{h}_{(\mu )}^{(p\pm )}\mid \mathfrak{h}_{(\nu )}^{(p\pm )}>_{%
\mathfrak{H}^{(\text{\textsc{H}}p\pm )}}.  \label{f9}
\end{eqnarray}%
We should notice that (\ref{6Lin}) and (\ref{f9}) agree with the fundamental 
$SU(2,2)$-law%
\begin{equation}
g_{\mu \nu }^{\prime \pm }=<e_{(\mu )}^{\prime \pm }\mid e_{(\nu )}^{\prime
\pm }>_{g^{\prime \pm }}=<e_{(\mu )}^{(p\pm )}\mid e_{(\nu )}^{(p\pm
)}>_{g^{\pm }}=g_{\mu \nu }^{\pm }.  \label{f10}
\end{equation}%
Additionally, we can see that the prescription (\ref{f6}) lies outside the
intersection (\ref{2}), that is to say,%
\begin{equation}
(u_{Dyn}^{\pm \bigstar }\text{\textsc{H}}^{(p\pm )})^{\bigstar }u_{Dyn}^{\pm
\bigstar }\text{\textsc{H}}^{(p\pm )}=\text{\textsc{H}}^{\bigstar (p\pm
)}u_{Dyn}^{\pm }u_{Dyn}^{\pm \bigstar }\text{\textsc{H}}^{(p\pm )}=\text{%
\textsc{H}}^{\bigstar (p\pm )}\text{\textsc{H}}^{(p\pm )}\neq I^{\pm
}(p^{\pm }),  \label{f}
\end{equation}%
with $I^{\pm }(p^{\pm })$ denoting the identity operator of $\mathfrak{C}%
^{\pm }(p^{\pm })\mathfrak{.}$ The spectra (\ref{spec1}) and (\ref{spec1lin}%
) are required to be invariant whence the helicities, spins and
polarizations of $(p^{+},p^{-})$ must correspondingly undergo the
transformations%
\begin{equation}
h^{\prime (p\pm )}=h^{(p\pm )},\text{ }\Sigma ^{\prime (f\pm )}=u_{Dyn}^{\pm
}\Sigma ^{(f\pm )}u_{Dyn}^{\pm \bigstar },\text{ }\Pi ^{\prime (b\pm
)}=u_{Dyn}^{\pm }\Pi ^{(b\pm )}u_{Dyn}^{\pm \bigstar }.  \label{f15}
\end{equation}%
Fitting together (\ref{f6}) and (\ref{f15}) thus supplies the behaviours of
the prescriptions (\ref{6}) and (\ref{8}), while Eq. (\ref{5Lin}) appears to
behave invariantly as shown by\footnote{%
Equation (\ref{g}) involves implementing (\ref{4}) together with the usual
definition of pseudo-unitary operators in the representation space: $%
uu^{\bigstar }=$ identity $\Leftrightarrow ugu^{\dagger }=g.$}%
\begin{eqnarray}
\mathfrak{H}^{\prime (\text{\textsc{H}}^{\prime }p\pm )} &=&\text{\textsc{H}}%
^{\prime (p\pm )\dag }g^{\prime \pm }\text{\textsc{H}}^{\prime (p\pm
)}=(u_{Dyn}^{\pm \bigstar }\text{\textsc{H}}^{(p\pm )})^{\dag }g^{\pm
}u_{Dyn}^{\pm \bigstar }\text{\textsc{H}}^{(p\pm )}  \notag \\
&=&\text{\textsc{H}}^{(p\pm )\dag }u_{Dyn}^{\pm }g^{\pm }u_{Dyn}^{\pm \dag }%
\text{\textsc{H}}^{(p\pm )}=\text{\textsc{H}}^{(p\pm )\dag }g^{\pm }\text{%
\textsc{H}}^{(p\pm )}=\mathfrak{H}^{(\text{\textsc{H}}p\pm )}.  \label{g}
\end{eqnarray}

When expressed in any admissible\ basis, covariant states are naturally
invariant in contradistinction to the case where the state superpositions
are posed in terms of $\mathfrak{h}$-bases. As amplitudes of states should
be preserved under local transformations like (\ref{3}), we can infer that,
under (\ref{f5}), the amplitudes of any outer-product states must behave as
trivial juxtapositions of conjugate pieces of the types\footnote{%
Henceforth, it shall sometimes be necessary to make use of the adjoint
version $g_{\pm }^{\ast \lambda \sigma }$ of $g_{\mu \nu }^{\pm }$ as well
as of some adjoint matrix relations. The adjoint description of the
representation space is given in Sections 2 and 7 of Ref. [7].}%
\begin{equation}
A_{\pm }^{\prime \lambda }=A_{\pm }^{\mu }u_{\mu \sigma }^{\pm \bigstar
}{}(Dyn)g_{\pm }^{\ast \sigma \lambda },\text{ }\overline{A_{\pm }^{\prime
\lambda }}=g_{\pm }^{\ast \lambda \sigma }u_{\sigma \mu }^{\pm }{}(Dyn)%
\overline{A_{\pm }^{\mu }}.  \label{f16}
\end{equation}%
It follows that the behaviour of a state like (\ref{16}) for some particle $%
p^{\pm },$ is described by%
\begin{equation}
<S^{\prime \pm }(p^{\pm })\mid =S_{\pm }^{\mu }(p^{\pm })u_{\mu \sigma
}^{\pm \bigstar }{}(Dyn)g_{\pm }^{\ast \sigma \lambda }<\mathfrak{h}%
_{(\lambda )}^{(p\pm )}\mid ,  \label{f30}
\end{equation}%
whereas the normalization condition (\ref{a3}) behaves invariantly as can be
directly seen by means of the calculation%
\begin{eqnarray}
&<&B^{\prime \pm }\mid \mathfrak{H}^{\prime (\text{\textsc{HB}}^{\prime }\pm
)}\parallel B^{\prime \pm }>_{\mathfrak{H}^{(\text{\textsc{HB}}^{\prime }\pm
)}}  \notag \\
&=&B_{\pm }^{\mu }(p^{\pm })u_{\mu \sigma }^{\pm \bigstar }{}(Dyn)g_{\pm
}^{\ast \sigma \lambda }\Delta _{\lambda \rho }^{\pm }g_{\pm }^{\ast \rho
\tau }u_{\tau \theta }^{\pm }{}(Dyn)\overline{B_{\pm }^{\theta }}  \notag \\
&=&B_{\pm }^{\mu }(p^{\pm })u_{\mu \sigma }^{\dag \pm }{}(Dyn)\Delta _{\pm
}^{\ast \sigma \tau }u_{\tau \theta }^{\pm }{}(Dyn)\overline{B_{\pm
}^{\theta }}  \notag \\
&=&B_{\pm }^{\mu }(p^{\pm })\Delta _{\mu \theta }^{\pm }\overline{B_{\pm
}^{\theta }}=<B^{\pm }\mid \mathfrak{H}^{(\text{\textsc{HB}}\pm )}\parallel
B^{\pm }>_{\mathfrak{H}^{(\text{\textsc{HB}}\pm )}},  \label{c1}
\end{eqnarray}%
which takes up (\ref{cond1}) along with (\ref{f4}) and the well-known
relations [7,10]%
\begin{equation}
g_{\pm }^{\ast \sigma \lambda }\Delta _{\lambda \rho }^{\pm }g_{\pm }^{\ast
\rho \tau }=\Delta _{\pm }^{\ast \sigma \tau },\text{ }u_{\mu \sigma }^{\dag
\pm }{}(Dyn)\Delta _{\pm }^{\ast \sigma \tau }u_{\tau \theta }^{\pm
}{}(Dyn)=\Delta _{\mu \theta }^{\pm }.  \label{r}
\end{equation}

The behaviours of the states (\ref{17}), (\ref{proc7}) and (\ref{proc4}) are
easily attainable because of the invariance of the states (\ref{4}).
Provided that the amplitudes of those states amount to nothing else but
particular coefficients of outer-product expansions, we may sort the
behaviours of such states by adequately juxtaposing patterns like (\ref{f16}%
). For instance, for the process that involves (\ref{17}), we have the
partial correlations%
\begin{eqnarray}
c_{+-}^{\prime 00} &<&0^{\prime (F+)}0^{\prime (F-)}\mid   \notag \\
&=&F_{+}^{\mu }u_{\mu \sigma }^{+\bigstar }{}(Dyn)g_{+}^{\ast \sigma
0}F_{-}^{\rho }u_{\rho \tau }^{-\bigstar }{}(Dyn)g_{-}^{\ast \tau
0}<0^{(F+)}0^{(F-)}\mid   \label{core1}
\end{eqnarray}%
and%
\begin{eqnarray}
c_{+-}^{\prime 11} &<&1^{\prime (F+)}1^{\prime (F-)}\mid   \notag \\
&=&F_{+}^{\mu }u_{\mu \sigma }^{+\bigstar }{}(Dyn)g_{+}^{\ast \sigma
1}F_{-}^{\rho }u_{\rho \tau }^{-\bigstar }{}(Dyn)g_{-}^{\ast \tau
1}<1^{(F+)}1^{(F-)}\mid .  \label{core2}
\end{eqnarray}

Equation (\ref{92}) as well as the normalization conditions for (\ref{proc7}%
) and (\ref{proc4}) all bear invariance. This can be brought forward by
making use of the above partial correlations and carrying out calculations
similar to that which led to (\ref{c1}). Creation and annihilation mappings
like the ones introduced by Eqs. (\ref{93}) and (\ref{proc21}), must be
regarded as invariant entities such that any statistical weights like (\ref%
{p1}), (\ref{p2}), (\ref{p4}) and (\ref{p5}) have to come into play as
dynamical invariants. Likewise, any entanglement operator like that defined
as Eq. (\ref{ent1}), should be related to an invariant physical property of
an annihilation process.

\section{Concluding Remarks}

We have suggested that the framework constructed in Ref. [7] should afford a
purely mechanical formulation of creation-annihilation processes wherein a
spontaneous formation of entangled states occurs. A somewhat important
result that has arisen from our work, is related to the obtainment of some
symbolic rules which make it feasible to keep track systematically in any
spacetime frame of helicity orientations for any process. In our view, any
creation-annihilation process is deemed to obey such rules because of the
elementary character that had been at the outset attributed to $\mathfrak{C}%
(p^{+},p^{-}).$ We believe that the applicability of the statistical
relationships carrying the $w$-weights we have associated to both types of
processes, could be verified experimentally. Whence, it can be said that a
physical meaning may perhaps be assigned to such relationships.

Remarkably enough, an opposite character has been attached to the helicities
of incoming and outgoing particles involved in WS-processes having two
outcomes. Nonetheless, we can expect that the incoming fermions involved in
annihilation processes having three outgoing bosons or antibosons, must
always carry either left-left or right-right helicities, with the entangled
bosonic outer-product kernels thus looking like%
\begin{eqnarray*}
&<&0^{(b\pm )}0^{(B\pm )}1^{(\text{\textsc{B}}\pm )}\mid ,\text{ }<0^{(b\pm
)}1^{(B\pm )}0^{(\text{\textsc{B}}\pm )}\mid , \\
&<&1^{(b\pm )}0^{(B\pm )}1^{(\text{\textsc{B}}\pm )}\mid ,\text{ }<1^{(b\pm
)}1^{(B\pm )}0^{(\text{\textsc{B}}\pm )}\mid , \\
&<&0^{(b\pm )}1^{(B\pm )}1^{(\text{\textsc{B}}\pm )}\mid ,\text{ }<1^{(b\pm
)}0^{(B\pm )}0^{(\text{\textsc{B}}\pm )}\mid .
\end{eqnarray*}

It is worth pointing out that if antibosons had been taken instead of bosons
in the process considered in Section 4, then we could get the corresponding
outgoing superposition by applying to Eq. (\ref{proc7}) the combined
replacements%
\begin{equation*}
b^{+}\rightarrow b^{-},\text{ }B^{+}\rightarrow B^{-},\text{ }\beta
_{++}^{01}\rightarrow \beta _{--}^{01},\text{ }\beta _{++}^{10}\rightarrow
\beta _{--}^{10}.
\end{equation*}%
In a WS-experiment having a boson and an antiboson as outcomes, the
entangled outgoing state would read%
\begin{equation*}
<\text{ent }b^{+}B^{-}\mid =\beta _{+-}^{00}<0^{(b+)}1^{(B-)}\mid +\text{ }%
\beta _{+-}^{11}<1^{(b+)}0^{(B-)}\mid ,
\end{equation*}%
but the incoming fermionic pair should once again make out the
configurations left-right and right-left just as in the case of (\ref{proc4}%
). Hence, in a horizontal-vertical formulation of annihilation processes
involving two outgoing bosons or antibosons, the spectral preservation
ensured locally by Eq. (\ref{3}) would tell us that the polarizations of the
outcomes should in either case bear orthogonality, in agreement with the
Dirac prediction mentioned in Section 1.

Our principal attitude at this stage has been to assume that entanglements
are formed spontaneously in any creation-annihilation process. We claim that
they must constitute a natural mechanical feature of any fundamental process.

\end{document}